# Lessons learnt from ITER safety & licensing for DEMO and future nuclear fusion facilities

Neill Taylor[a], Pierre Cortes[b]

[a]*EURATOM/CCFE Fusion Association, Culham Science Centre, Abingdon, Oxfordshire OX14 3DB, U.K.*
[b]*ITER Organization, Route de Vinon sur Verdon, 13115 St. Paul lez Durance, France*

One of the strong motivations for pursuing the development of fusion energy is its potentially low environmental impact and very good safety performance. But this safety and environmental potential can only be fully realized by careful design choices. For DEMO and other fusion facilities that will require nuclear licensing, S&E objectives and criteria should be set at an early stage and taken into account when choosing basic design options and throughout the design process.

Studies in recent decades of the safety of fusion power plant concepts give a useful basis on which to build the S&E approach and to assess the impact of design choices. The experience of licensing ITER is of particular value, even though there are some important differences between ITER and DEMO. The ITER project has developed a safety case, produced a preliminary safety report and had it examined by the French nuclear safety authorities, leading to the licence to construct the facility. The key technical issues that arose during this process are recalled, particularly those that may also have an impact on DEMO safety. These include issues related to postulated accident scenarios, environmental releases during operation, occupational radiation exposure, and radioactive waste.

## 1. Introduction

The ITER fusion device, currently under construction in the south of France, is the first nuclear fusion facility for which a full safety case has been prepared and subjected to the scrutiny of a nuclear regulator. The case was presented in the ITER safety files, in particular the preliminary safety report (*Rapport Préliminaire de Sûreté, RPrS*) [1,2] and examined by the French nuclear safety authorities (*Autorité de Sûreté Nucléaire, ASN*) and their technical advisors (*Institut de Radioprotection et de Sûreté Nucléaire, IRSN*) [3,4]. The safety files were also submitted to a public enquiry and assessed by the French environmental agency. The outcome of this stage of the licensing process was the granting by the French government, in November 2012, of the decree which authorizes the construction of the ITER nuclear facility.

This process ensured that the ITER design achieves the high standard of nuclear safety expected by the authorities and by international nuclear safety guidelines. The documented safety analysis has shown that the operation of ITER will meet its goals for minimization of the likelihood and consequences of potential accidents, with extremely low environmental impact during normal operation including maintenance activities, and that generated radioactive waste is manageable and reduced to the maximum extent possible. In the course of this safety and licensing process many lessons have been learned. Some of these are of direct relevance to DEMO or other future nuclear fusion facilities, and may inform considerations of design options from the very beginning of the conceptual design of these facilities.

In this paper some of the key technical issues that have been addressed are outlined, particularly where the issue may also be of relevance to DEMO (note that the term "DEMO" is used generically in the text, and refers also to other possible fusion facilities that may be conceived to come between ITER and a commercial power plant, such as a Component Test Facility or a Fusion Nuclear Science Facility).

## 2. Differences between ITER and DEMO

Before considering those aspects of ITER safety that may have relevance to DEMO, it is useful to recall the essential differences between the two projects. Some of these are listed in table 1. They may imply that some aspects of the safety approach employed for ITER may not be transferable to DEMO. However, many of the issues encountered in the safety and licensing of ITER are certainly of potential relevance to DEMO, including those described in the following sections.

## 3. Safety objectives and safety functions

The top-level safety objectives for ITER are based on international guidelines and are similar to those adopted by any nuclear facility, with the addition that the advantageous safety characteristics of fusion should lead to limited consequences of any accident. They are:

- to protect workers, the public and the environment from harm;
- to ensure in normal operation that exposure to hazards within the premises and due to release of hazardous material from the premises is controlled, kept below prescribed limits and minimized to be as low as reasonably achievable;

_____________________________________________
*author's email: neill.taylor@ccfe.ac.uk*

Table 1. Essential ITER/DEMO differences

| ITER | DEMO |
|---|---|
| Experimental device with physics and technology missions | Nearer to a commercial power plant, but with some development missions |
| 400s pulses (some longer at lower power), long dwell time | Long pulse, quasi-steady state |
| Experimental campaigns. Outages for maintenance and component replacements | Maximize availability |
| Large number of diagnostics | Only those required for operation |
| Multiple heating and current drive systems | Fewer |
| Large design margins, necessitated by uncertainties and lack of fully appropriate design codes | With ITER (and other) experience, design could have smaller uncertainties |
| Cooling system optimized for minimum stresses and sized for modest heat rejection | Cooling system optimized for electricity generation efficiency (e.g. higher temperature) |
| Test blanket modules introduce range of diverse concepts | Single blanket concept |
| Unique one-off design optimized for experimental goals within cost constraints | Move towards design choices suitable for series production |
| No tritium breeding requirement (except very small quantity in TBMs) | Tritium breeding needed for self-sufficiency |
| Conventional 316 stainless steel structure | Novel low activation materials as structure (at least for some components) |
| Very modest lifetime neutron fluence, low dpa and He production | High fluence, significant materials damage |
| Licensing as experimental facility allows some credit for experimental nature (e.g. no dependence of safety on plasma behavior) | Stricter approach may be necessary to avoid large design margins |
| During conceptual design (including "EDA"), licensing in any ITER member country had to be possible | Fewer constraints |

- to ensure that the likelihood of accidents is minimized and that their consequences are bounded;
- to ensure that the consequences of more frequent incidents, if any, are minor;
- to demonstrate that the favourable safety characteristics of fusion permit a safety approach that limits the hazards from accidents such that in any event there is no need for public evacuation on technical grounds;
- to minimize radioactive waste hazards and volumes and ensure that they are as low as reasonably achievable.

Similar objectives should be appropriate for DEMO, and are consistent with those adopted for earlier studies of the safety of fusion power plant, such as the European Power Plant Conceptual Study (PPCS) [5]. The goal of ensuring that public doses following an accident could never reach the level at which evacuation would be triggered has been a constant feature of such studies. The radioactive waste target has often been stated as ensuring that waste is not a burden on future generations, effectively requiring that all active material could be cleared or recycled within a period of up to 100 years.

In pursuing the above objectives, just two safety functions are defined for ITER:

- Confinement of radioactive material;
- Limitation of exposure to ionizing radiation.

These are also suitable safety functions for DEMO, with confinement preventing releases to the environment during normal operation and maintenance as well as in accident situations. In ITER there are also several non-safety supporting functions which enable the above two to be achieved in all situations. They include fire protection and prevention of explosions, and decay heat removal (see sections 5.2 and 5.3).

## 4. Confinement

The strategy for confinement in ITER is that every significant inventory of radioactive material is protected by two confinement systems, each comprising one or more physical or functional barriers. Broadly, the first confinement system prevents mobilization within the facility, and thereby protects workers, and the second confinement system prevents release to the environment in the event that the first confinement has failed. A similar strategy would be appropriate for DEMO and other fusion nuclear facilities, but care must be taken with the choice of the physical barriers that are to provide the confinement function.

In ITER, for the in-vessel inventory of tritium and activated erosion dust (assumed to be up to 1 kg and 1000 kg respectively) the first confinement system is the vacuum vessel (VV) and its extensions. The vessel is a robust double-walled chamber designed to resist, with a large margin, all foreseen loads including electromagnetic loads induced by the largest expected plasma vertical displacement event. But the VV "extensions" include not only the 45 ports but also hundreds of penetrations through the vacuum boundary for cooling systems, heating and current drive systems, diagnostics systems, in-vessel coil feeders, etc. All of these participate in providing the first confinement barrier and are therefore of highest safety importance class.

The situation with VV penetration systems is complicated by the firm policy of not crediting any in-vessel component with a safety function. In ITER this is essential because the experimental nature of these components implies that they are prone to failure and their reliability in accidental situations could not be guaranteed. In safety analyses it is generally assumed that there is a first wall or divertor failure in any relevant accident scenario. This means that, for water cooling circuits, for example, the section inside the VV has to be assumed to be breached, meaning that the ex-vessel part, at least up until safety-class isolation valves, is part of the first confinement system and must be reliably leak-free in all situations including earthquake. This results in a first confinement barrier that comprises many tens of km of pipework, ducts, waveguides etc. and hundreds of isolation valves and windows including non-metallic windows for which a qualification is needed.

For electricity-producing fusion power plant of the future, in-vessel components must be highly reliable and damaging plasma disruptions must be eliminated. At that stage it will be possible to credit these components with some safety function, if required. DEMO and other mid-term nuclear fusion facilities are at some intermediate stage. In-vessel components such as the divertor and first wall/blankets will still be at a developmental stage, but some of the parts inside the VV, such as sections of cooling circuit, could possibly be credited with a confinement function in order to avoid the extensive and complex ex-vessel parts that complicate the first confinement barrier in ITER.

## 5. Accidents

### 5.1 Accident analyses

The chief concern for the consequences of abnormal events is the potential to damage, degrade or bypass the confinement systems. Such an accident could lead to mobilization of radioactive material within the facility, and consequent exposure of personnel, and ultimately to the release to the environment of some part of the radioactive inventory and possible public radiation dose. In the ITER safety case, extensive analysis of postulated accident scenarios have been performed [6], and have shown that the worst consequences in terms of public doses are minor.

Accident analyses have similarly been performed in the past as part of studies of fusion power plant concepts [7] and have reached similar conclusions for the range of event scenarios studied. For DEMO, it will be necessary to perform more detailed analyses as designs are developed and, as in the ITER approach, these will indicate what supporting safety functions are required and what systems need to be credited in order to maintain low consequences of any conceivable event.

In ITER, a potential accident sequence arises from the possibility of a break in a cryoline or cryogenically-cooled component and release into the building of large quantities of helium cryogen. This could lead to over-pressure and under-temperature stresses and a challenge to a confinement system. The potential is particularly of concern if the leak is into a room containing a radioactive inventory. This hazard will exist in any facility with a large cryogenic and cryo-distribution system. Analyses must be performed to ensure that the stresses can be accommodated and/or that adequate pressure-relief provisions are in place.

As has been done for ITER, a full range of external events should be considered as accident initiators, including combinations of events. Even for extremely improbable combinations of events, it is likely to be required to show that consequences are limited and that there is no "cliff-edge" effect.

### 5.2 Fire and explosion

Following the experience of ITER, particular attention should be afforded to the potential for fire and explosion events involving hydrogen. Substantial inventories of hydrogen isotopes will be present in parts of the ITER plant, and it might be expected that in DEMO these could be even higher, owing to the higher throughput of deuterium and tritium fuels.

In ITER, the hydrogen explosion risks fall into two types. The first relates to the risk of a deflagration of hydrogen within building rooms housing systems of the fuel cycle, including vacuum pumping, fuel processing, fuel storage, and fuel injection systems. A flammable or explosible hydrogen/air mixture could result from an accidental leak of hydrogen, deuterium or tritium from fuel cycle equipment. A similar risk may exist elsewhere, for example in the hot cell where tritium is recovered from dust or components that have been removed from the VV. A deflagration or detonation of such a mixture could challenge the integrity of building walls and slabs that are providing part of the second confinement system. If the explosion involves tritium, then the radioactive source term for release through a damaged confinement is already available, i.e. the event involves the breach of both confinement systems. This event can be avoided or mitigated by a combination of measures:

- fire detection and suppression systems,
- air mixing systems to prevent local hydrogen concentrations above the flammable limit,
- establishment of a fire sectorization with limited tritium inventory in each sector (where possible), to restrict the amount releasable in a single fire,
- design building walls and slabs to resist the maximum overpressure foreseen in postulated explosion events – this requires extensive analysis.

Similar provisions to these must be taken in DEMO, where the risks will be similar. In particular attention must be paid to the fire sectoring and overpressure requirements at an early stage of building layout design.

The ignition energy for hydrogen/air is very low, so avoiding ignition sources is not practicable.

The second area of concern for hydrogen explosions is in-vessel. Here there is a significant inventory of deuterium and tritium trapped in plasma-facing components and absorbed in dust, as well as stored on cryopump panels awaiting regeneration. This inventory could be augmented in certain accident scenarios, see below. It is credible to postulate an accident that involves an ingress of air into the VV, a "Loss of vacuum accident" (LOVA), leading to mobilization of a part of this inventory and the formation of a hydrogen/air mixture above the flammable limit. The small energy required to ignite the mixture is readily available from hot surfaces.

In ITER, an additional hydrogen inventory may be generated in the case of an ingress of coolant into the vessel following damage of a first wall or divertor. A chemical reaction can be foreseen between steam and beryllium, either on the first wall or as dust, which may have high porosity. This reaction produces hydrogen and is exothermic. In another facility such as DEMO, which is expected to use tungsten in place of beryllium as plasma-facing material, this source of hydrogen could be greatly reduced. If a coolant other than water is used, it could be eliminated.

It is not clear if the timescales for the mobilization and generation of hydrogen isotopes from these various sources are similar, and therefore whether they should be considered in combination for an explosion scenario. But if a hydrogen explosion does occur, it has the potential to create a transient overpressure that exceeds the design pressure of the vessel and other parts of the vacuum boundary (0.2 MPa abs. in ITER). Furthermore it can be postulated that even a relatively small hydrogen explosion could mobilize collected beryllium dust and ignite it in a dust cloud explosion that could also create an overpressure challenging the confinement barrier. Again, for DEMO, it may be possible to reduce this potential aggravation by choosing a plasma-facing material other than beryllium.

If the possibility of hydrogen and dust explosions cannot be eliminated by choice of materials, it may be feasible to minimize the associated risk by active design measures. For ITER, consideration is being given to installing a mitigation solution based either on igniters within the vessel, which could ignite a hydrogen/air mixture as soon as the lower flammability limit is reached and resulting in a harmless low-pressure combustion, or a rapid injection into the vessel of an inert gas, triggered by the detection of air ingress into the vessel (a LOVA) [8]. The igniters solution could be hot-wire filaments with battery-backed power supply, constantly hot while the vessel is under vacuum and thus highly reliable. The locations for these igniters must be chosen with the aid of CFD modeling to maximize their effectiveness. For the gas injection solution, the location of the injection points must be chosen with care to avoid the dynamic effect of the gas injection causing a local concentration of hydrogen.

A further option for DEMO, not considered for ITER, would be the filling of volumes surrounding the VV with an inert gas, to avoid any possibility of air ingress. This would have to include any rooms or cells where there are penetration lines that extend the vacuum boundary, and for this reason may not be practical.

### 5.3 Decay heat removal

For ITER the removal of decay heat after plasma shutdown is regarded as a supporting function, not a primary safety function. Failure to remove decay heat does not in itself lead to any safety consequence, as temperatures rise only very slowly and do not reach the level at which any structural degradation could be expected. For a fusion plant operating at substantially higher power than ITER's 500 MW and close to steady-state or high duty cycle pulses, the removal of decay heat could be classed as an additional safety function.

The level of decay heat in ITER is low, starting at a plant total of around 10 MW at shutdown and falling below 1 MW within one day, corresponding to an average heat density in all in-vessel components of less than 1 W/kg. For DEMO, the low activation materials that may be used for plasma-facing components produce lower decay heat than the 316 stainless steel in ITER, but this is compensated for by higher plasma power and much higher duty cycle. In conceptual power plant studies [5], the total decay heat densities in in-vessel components based on Eurofer low-activation martensitic-ferritic steel are typically around 10 W/kg after 1 day, and a DEMO using this structural material could be expected to be similar. This order of magnitude increase compared with ITER may be enough to make decay heat a potential hazard in its own right, and its removal a safety function.

The removal of decay heat in ITER is done via the VV coolant loop, which, even with a slow rate of circulation, can reject the heat not only of the vessel itself but also of all in-vessel components as their heat is conducted and radiated out to the vessel walls. This low coolant flow can be maintained by a small pump powered by the emergency diesel generators in the event of a loss of power. If this too fails, studies have shown that the introduction of a fluid, such as air, into the cryostat volume is sufficient to allow a passive removal of the heat by convection and conduction to the cold magnet structures and to the cryostat wall and the atmosphere outside [2]. A similar approach may be suitable for DEMO, and a passive means of ultimate heat removal is to be favoured.

### 5.4 Control of tritium inventories

As noted in section 5.2, maintaining a restriction on the quantity of tritium that may be vulnerable to release in an accident scenario such as a fire is a key provision to limit the severity of the consequences. This implies a strict control on the inventory of tritium in each process and storage system, as well as within each room (or fire sector), including that of trapped tritium such as inside the VV. Tritium accountancy in a complex system can be challenging; at ITER it will be based on maintaining a number of mass balance areas [9].

For DEMO the implementation of a tritium accountancy system is further complicated by the additional tritium produced in the breeding blankets and returned to the fuel cycle by an extraction system. The increased throughput of tritium via fuelling and pumping systems may also increase uncertainties.

## 6. Environmental releases in normal operation

The potential for routine releases of radioactive material in gaseous or liquid form has been assessed in earlier studies of fusion power plant concepts [5], with the conclusion that emissions would be very low, with a maximum public dose impact around 1 μSv/year. This may now appear to be an under-estimate. For ITER a more extensive study was carried out, identifying a wider range of potential contributions to releases. For gaseous and aerosol releases, it was recognized that these may come from a number of sources:

- tritium leaks from fuel cycle equipment during normal operation and maintenance;
- outgassing of tritium from components removed from the vessel during their storage or processing in the hot cell including storage while waiting for disposal;
- leaks from the tritium recovery station used to remove tritium from dust that has been collected in the vessel or from solid components before disposal;
- outgassing of tritium and suspended dust particles via the vessel vent system during periods that the VV is vented for maintenance.

The earlier power plant studies only considered in detail the first of these sources. For DEMO it will be essential to analyze them all and to implement strategies to minimize environmental releases. As at ITER, this can be achieved by filtered venting and detritiation systems that maintain a sub-atmospheric pressure in the volumes into which the leaks may occur. There may be very small additional tritium releases via permeation into rooms *not* served by detritiation systems, or into cooling water circuits and released at the cooling towers; this should be assessed. Detritiation systems must be highly reliable and maintain a specified efficiency in all normal and abnormal situations. Additional measures at ITER to reduce tritium releases include the baking of the divertor at 350°C before any venting and opening of the vessel ports. This is the region in which dust is expected to accumulate, and the baking promotes outgassing of absorbed tritium which can then be returned to the fuel cycle via the normal pumping system.

In a water-cooled tokamak such as ITER, the potential for liquid releases to the environment arises principally from leaks from the primary cooling circuit. This can be expected to contain an inventory of activated corrosion products as well as tritium through permeation. Other sources of contaminated water may arise from processes that employ water, from water used in cleaning and decontamination procedures, from condensation in contaminated atmospheres such as the hot cell, and from test samples in laboratories. Secondary water coolant loops may also become low-level contaminated, even with a gaseous primary, particularly by tritium permeation in the heat exchanger or by small heat exchanger tube leaks.

The outcome of this is that the plant may contain water with a wide range of contamination levels. At ITER, water with relatively high tritium content can be sent to the Water Detritiation System within the tritium plant, where the tritium is recovered. Water with a very low level of tritium content can be discharged, after monitoring, as industrial waste. But there is a range of tritium content in between these two levels at which there is no on-site destination. For ITER, it is not expected to generate water contaminated in this range, but if it did occur it may have to be transported off site to another nuclear facility that could treat it. For DEMO, it is important to ensure that the full range of anticipated contamination levels is identified and treatment or disposal paths implemented for all parts of it.

The estimation of public doses as a consequence of environmental release of tritium depends on a prediction of the dispersion of HT and HTO, the biological hazard potential if the tritium is inhaled or ingested, and in the latter case the pathways for tritium to reach the food chain. This includes the formation of organically bound tritium in vegetable matter. These processes are well represented in dispersion and dose modeling [10], but there may be scope for further improvement.

## 7. Solid radioactive waste

Safety and environmental studies of conceptual fusion power plant [5] have shown that by the use of low activation materials for components close to the plasma it is possible to restrict activation to the level that most – perhaps all – of the activated material will decay within a century to the level at which it could be cleared (i.e. recycled with no restriction on its use) or recycled within the nuclear industry. The criteria for permitting clearance or recycling vary from country to country [11] and in some studies the limits for the recyclability of material have been set in terms of radiological criteria somewhat arbitrarily. It will be essential to develop viable industrial processes for recycling procedures, and at the same time establish realistic criteria. Nevertheless the studies give confidence that it will be possible to avoid an accumulation of radioactive waste that could be a burden on future generations.

ITER is being fabricated of austeninic stainless steel, with a few specific impurities controlled to minimize long-term activation. This steel will reach a higher level of activation, per incident neutron, than the low activation materials under development for a power plant. However the integrated lifetime neutron fluence in ITER, 0.3 MW.yr/m$^2$, is very much lower than that expected of a typical power plant, which could be up to 50 MW.yr/m$^2$ over a 20-year plant life. The combination of these competing effects is that the level of activation of material in a power plant, and hence its classification, could be similar to that of ITER. DEMO is likely to be in an intermediate position, with fluence somewhat less than a power plant, but maybe with less than full deployment of low activation materials. Thus again, the level of activation of waste could be similar to ITER, and the quantity too for concepts that are of a similar major radius to that of ITER.

An issue that arose during the development of the waste management strategy for ITER is that material that is both activated and contaminated with tritium does not at present have a disposal path within the French radioactive waste repositories. To resolve this an interim waste store is planned to allow for several half-lives decay of tritium before its final disposal [12]. This issue is likely to be also applicable to DEMO and other future facilities, unless an effective detritiation technique for solid radioactive waste is developed and implemented.

## 8. Occupational Radiation Exposure

In a nuclear fusion facility such as ITER or DEMO the requirement to keep personnel doses As Low As Reasonably Achievable is achieved by a rather conventional approach:

- Maximize the use of remote handling to avoid human intervention in active areas;
- Minimize shutdown gamma dose rates by design of components and choice of low-activation materials;
- Optimize maintenance procedures to reduce the duration of human access to areas of high dose rate;
- Provide adequate shielding to keep dose rates low;
- Implement an access control system to impose a radiological zoning scheme.

At ITER an optimization is applied to the design to reduce the potential for occupational doses to the maximum extent realistically possible [13].

Access to the tokamak during plasma operation is prohibited including all rooms containing parts of the primary water cooling circuit, due to intense high-energy gamma radiation from the decay of $^{16}$N. This nuclide is generated in water by the reaction $^{16}$O $(n,p)$ $^{16}$N and is therefore unavoidable. As this reaction has a neutron energy threshold of 10.5 MeV, it is far more important in D-T fusion devices compared with other nuclear facilities such as fission reactors. Although $^{16}$N has a half-life of only 7.1 s, it is sufficient to make all parts of a water cooling circuit a prominent source of radiation during operation, including those parts lying outside the bioshield. Of course, for DEMO this problem could be excluded by choosing a coolant other than water.

## 9. Conclusions

The experience of the ITER project in developing a safety approach, implementing it in a safety design, and performing safety analyses under the scrutiny of a nuclear regulator is of relevance to DEMO and other future nuclear fusion facilities. Although there are some important differences between ITER and DEMO, many of the issues encountered during the development of the ITER safety case and its defence before the French nuclear safety authorities are of potential importance to DEMO. Some of these may be eliminated or ameliorated by fundamental design choices, and so should be taken into account from the very beginning of DEMO design activities.

Finally it should be noted that the scope of this paper has been limited to radiological hazards. Non-nuclear hazards and potential environmental impacts, for example associated with the use of beryllium, must also be addressed.

## Acknowledgments

This work was part-funded by the RCUK Energy Programme under grant EP/I501045 and the European Communities under the contract of Association between EURATOM and CCFE. To obtain further information on the data and models underlying this paper please contact PublicationsManager@ccfe.ac.uk. The views and opinions expressed herein do not necessarily reflect those of the European Commission nor those of the ITER Organization.

## References

[1] N. Taylor et. al., Preliminary safety analysis of ITER, Fusion Sci. Technol. 56 (2009) 573.
[2] N. Taylor et. al., Updated safety analysis of ITER, Fusion Eng. Des., 86 (2011) 619.
[3] N. Taylor et al., ITER safety and licensing update, Fusion Eng. Des. 87 (2012) 476-481
[4] N. Taylor, C. Alejaldre and P. Cortes, Progress in the Safety and Licensing of ITER, Fusion Sci. Technol. 64 (2013) 111-117.
[5] D. Maisonnier et. al., The European Power Plant Conceptual Study, Fusion Eng. Des., 75–79 (2005) 1173-1179.
[6] S. Reyes et. al., Updated Modeling of Postulated Accident Scenarios in ITER, Fusion Sci. Technol., 56 (2009) 789.
[7] I. Cook et. al., European Fusion Power Plant Studies, Fusion Sci. Technol. 47 (2005) 384-392.
[8] J. Xiao, J.R. Travis, W. Breitung and T. Jordan, Numerical analysis of hydrogen risk mitigation measures for support of ITER licensing, Fusion Eng. Des., 85 (2010) 205-214.
[9] M. Glugla et. al., The ITER Tritium Systems, Fusion Eng. Des. 82 (2007) 472–487.
[10] N.P. Taylor and W. Raskob, Updated accident consequence analyses for ITER at Cadarache, Fusion Sci. Technol. 52 (2007) 359-366.
[11] M. Zucchetti et. al., The feasibility of recycling and clearance of active materials from fusion power plants, J. Nucl. Mater. 367–370 B (2007) 1355-1360.
[12] J. Paméla, ITER tritiated waste management by the Host state, these proceedings.
[13] P. Cortes et. al., Optimization at the design phase of the potential impact of ITER on workers, the public and the environment, Fusion Eng. Des. 85 (2010) 2263–2267.